\documentclass[fleqn,11pt]{article}

\usepackage{amssymb,amsmath,color}

\usepackage{amsfonts,amssymb,cite}
\usepackage{graphicx}

\def\noi{\noindent}

\newcommand{\Title}[1]{\noi {{\Large\bf #1}}\\[1ex]}

\def\Aunames#1{\noi{\bf #1}}

\def\Addresses#1{\medskip\noi \protect
	\begin{description}\itemsep -3pt {\it #1} \end{description}}
\def\adr#1#2{\item[${}^{#1}$]{\it #2}}

\newcommand{\Abstract}[1]{\vskip 2mm \begin{center}
        \parbox{16.4cm}{\small\noi #1} \end{center}\medskip}

\def\email#1#2{\footnotetext[#1]{e-mail: #2}\addtocounter{footnote}{1}}

\def\nqq{\hspace*{-2em}}

\usepackage{color}

\def\Jl#1#2{#1 {\bf #2},\ }

\def\ApJ#1 {\Jl{Astroph. J.}{#1}}
\def\CQG#1 {\Jl{Class. Quantum Grav.}{#1}}
\def\DAN#1 {\Jl{Dokl. AN SSSR}{#1}}
\def\GC#1 {\Jl{Grav. Cosmol.}{#1}}
\def\GRG#1 {\Jl{Gen. Rel. Grav.}{#1}}
\def\IJMPD#1 {\Jl{Int. J. Mod. Phys. D}{#1}}
\def\JETF#1 {\Jl{Zh. Eksp. Teor. Fiz.}{#1}}
\def\JETP#1 {\Jl{Sov. Phys. JETP}{#1}}
\def\JHEP#1 {\Jl{JHEP}{#1}}
\def\JMP#1 {\Jl{J. Math. Phys.}{#1}}
\def\NPB#1 {\Jl{Nucl. Phys. B}{#1}}
\def\NP#1 {\Jl{Nucl. Phys.}{#1}}
\def\PLA#1 {\Jl{Phys. Lett. A}{#1}}
\def\PLB#1 {\Jl{Phys. Lett. B}{#1}}
\def\PRD#1 {\Jl{Phys. Rev. D}{#1}}
\def\PRL#1 {\Jl{Phys. Rev. Lett.}{#1}}

\def\lal{&&\nqq {}}

\def\beq{\begin{equation}}
\def\eeq{\end{equation}}
\def\bear{\begin{eqnarray}}
\def\bearr{\begin{eqnarray} \lal}
\def\ear{\end{eqnarray}}
\def\earn{\nonumber \end{eqnarray}}

\def\const{{\rm const}}

 \catcode`\@=11 \@addtoreset{equation}{section}\catcode`\@=12

\def\R{{\mathbb R}}
\def\seq{{\quad \Rightarrow \quad}}

\begin{document}
\twocolumn[


\Title{Photon spheres near  black holes in a model
       with anisotropic fluid   }

\Aunames{V. D. Ivashchuk$^{a,b,1}$, S. V. Bolokhov$^{a}$, F. B. Belissarova$^{c}$, N. Kydyrbay$^{c}$, A. N. Malybayev$^{c}$,  G. S. Nurbakova$^{c}$,    and  B. Zheng$^{a}$ } 

\Addresses{
\adr a {\small Peoples' Friendship University of Russia (RUDN University), 
             ul. Miklukho-Maklaya 6, Moscow 117198, Russia} \\         
\adr b {\small Center for Gravitation and Fundamental Metrology, SCAM Rostest,
            ul. Ozyornaya  46, Moscow 119361, Russia} \\
\adr c {\small 
    Al-Farabi Kazakh National University,  Al-Farabi avenue, 71, Almaty 050040, Kazakhstan } 
	}


\Abstract
{
The semi-review paper studies the null geodesics which appear
for black hole  solutions 
 in the gravitational $4d$ model  with anisotropic fluid. 
The equations of state for the fluid and solutions itselves depend upon integer 
parameter  $q = 1, 2, ...$:   $p_r = -\rho c^2 (2q-1)^{-1}, \quad p_t = - p_r$,
where $\rho$ is the mass density, $c$ is speed of light, 
$p_r$  and $p_t$  are  pressures in radial and transverse  
to radial directions, respectively.
The circular null geodesics are explored and the master equation for radius
$r_*$ of  photon sphere is outlined as well as the proposition on existence and 
uniqueness of the solution to master equation
obeying $r_* > r_h$, where  $r_h$ is horizon radius. The relations for spectrum of  quasinormal modes  for  a
 test  massless scalar field in the eikonal approximation are overviewed and 
 compared with cyclic frequencies of circular null geodesics.  The shadow angles are explored. 
 }
 ] 

\email 1 {ivashchuk@mail.ru}

{ 
\def\R{{\mathbb R}}

\section{Introduction}

Recent discovery of gravitational waves 
emitted from two black hole  merging  \cite{Abbott} 
 has strengthen an interest in studying  black holes which was  
inspired originally by first evidence of supermassive black hole in the center 
of our galaxy \cite{Ghez}. At the moment a lot of papers are
devoted to black holes in modified theories of gravity, for a review
see \cite{Vagn} and references therein. Here we consider 
one of such theories - the gravitational model with anisotropic fluid 
as a matter source.      

In this paper we continue our previous works \cite{DIM,BI}
 devoted to  black hole solutions in the model with anisotropic fluid. 
Here we deal mainly with  null geodesics which appear for black hole  solutions 
from Ref. \cite{DIM,BI}. 
The equations of state for the fluid and solutions itselves depend upon integer 
parameter  $q = 1, 2, 3, \dots$.
   
  Here we explore the  circular null geodesics with constant radius $r_*$ which
is the radius of photon  sphere. Parameters of photon spheres play a key role 
for numerous applications of black hole solutions:  
spectra of  quasinormal modes (QNM) in the eikonal approximation,  
 circular orbits for massive particles, black hole shadows etc, see 
 \cite{Vir,KZh,CGP,KSt,MBI,IMNT,BTIMNU}  and references therein.  

The paper has the following structure. 
In Section 2 we describe the model and  the black hole  solutions with anisotropic fluid. 
In Section 3 we explore null geodesics for the black hole  solutions under consideration.  
In Section 4 we  deal with  master equation for radii $r_*$ of a circular photon orbits 
in the background black hole metrics under consideration and 
 present a proposition on existence and uniquenes of the solution
 to the  master equation obeying $r_{*}  > r_h$, where  $r_h$ is horizon radius.
 We present explicit relations for finding  $r_{*}$, when $q = 1, 2, 3$ and $q = + \infty$. 
In Section 5 we present relations for shadow angles  and outline a  spectrum of 
 quasinormal modes for  a  test neutral massless scalar field in the eikonal approximation 
 for any $q = 1,2,...$.

\section{Black hole solutions}

Here we consider the solutions to Einstein equations
\begin{equation}
\label{b1.0}
R^{\mu}_{\nu} - \frac{1}{2} \delta^{\mu}_{\nu} R = \kappa T^{\mu}_{\nu},
\end{equation}
 where $\kappa = 8 \pi G /c^4$, $G$ is Newton gravitational constant and $c$ is speed of light. 

The solutions under consideration are defined on the four-dimensional manifold with topology
\begin{equation}
\label{b1.1}
  M = \mathbb{R}_{(\rm radial)}\times \mathbb{S}^{2}\times
\mathbb{R}_{(\rm time)}.
\end{equation}
Here the spherical coordinate system is used: $ (x^\mu) =(r, \theta, \phi, t)$  with signature $(+++,-)$.
The energy-momentum tensor of anisotropic fluid is taken as
\begin{equation}
\label{b1.2}
 (T^{\mu}_{\nu})={\rm diag}\left(p_r,\ p_t, p_t, \ -\rho c^2 \right),
\end{equation}
and the equations of state read
\begin{equation}
\label{b1.3}
  p_r = -\rho c^2 (2q-1)^{-1}, \qquad p_t = - p_r.
\end{equation}
Here $\rho$ is the mass density, $p_r$  and $p_t$  are  pressures in radial and orthogonal 
(to radial) directions, respectively.   

The parameter $q$ describes relations between the pressures and the mass density; $q>0$, $q\neq
1/2$. In the present paper, the parameter $q$ is taken to be a natural number to avoid the non-analytical
behaviour of the metric at the (would be) horizon.

The solution has the following form \cite{DIM}:

\begin{eqnarray} 
  \nonumber
  ds^2= g_{\mu \nu} dx^{\mu} dx^{\nu}   \\ 
    = (H(r))^{2/q}  
    \left[ \frac{dr^2}{1-\frac{2\mu}{r}}
 + r^2 d\Omega^2  \right.   \nonumber \\
  \left. -(H(r))^{-4/q}\left(1-\frac{2\mu}{r}\right) c^2 dt^2 \right],
  \label{b1.4}
\end{eqnarray}
\begin{equation}
 \label{b1.5}
 \kappa \rho c^2 = \frac{(2q-1)P(P+2\mu)(1-2\mu r^{-1})^{q-1}}{H(r)^{2+\frac{2}{q}}\; r^{4}},
\end{equation}
where the function $H(r)$ reads as follows:
\begin{equation}
 \label{b1.6}
 H(r) = 1+\frac{P}{2\mu}
 \left[1-\left(1-\frac{2\mu}{r}\right)^q \right].
\end{equation}
The metric on the unit 2-dimensional sphere $\mathbb{S}^2$ is denoted by
 $d\Omega^2  = d \theta^2 + \sin^2 \theta d \phi^2$
 ($0< \theta < \pi$, $0< \phi < 2 \pi$); parameters  $P,\mu > 0$ are arbitrary. 
 Originally we put $r > 2\mu = r_h$
 but the domain of definition of the metric may be extended \cite{BI}. 

The equations of motion (\ref{b1.0}) imply the following relation for the
scalar curvature
\begin{equation}
\label{b1.7}
R[g] = - \kappa T^{\mu}_{\mu} = \frac{2(q-1)}{2 q -1} \kappa \rho c^2.
\end{equation}
which may be used  for identifying the singularities of solutions
for $q = 2,3,4,\dots$ \cite{BI}.

\section{Physical parameters}

Here we present certain physical parameters
corresponding to our black hole solutions \cite{BI}.
Here we put for simplicity $c = \hbar = k_B = 1$.

The ADM gravitational mass for black hole under consideration
reads as follows
\begin{equation} 
 \label{b3.12}
 GM = \mu +  \frac{P}{q}.
 \end{equation}
For post-Newtonian parameters we get 
\begin{eqnarray} 
 \label{b3.13}
 \beta =  1 + \frac{q P (P + 2 \mu) }{2(GM)^2}, \\
 \label{b3.13a}
 \gamma = 1.
\end{eqnarray}

The Hawking temperature of the black hole is given by relation
\begin{equation} 
 \label{b3.H}
  T_H = \frac{1}{8 \pi \mu}   \left(1 + \frac{P}{2\mu} \right)^{- 2/q},
\end{equation} 
while the Bekenstein-Hawking (area) entropy  corresponding to
the horizon at $r = 2 \mu$  reads
\begin{equation} 
 \label{b3.S}
  S_{BH} = \frac{4 \pi \mu^2}{G}   \left(1 + \frac{P}{2\mu} \right)^{2/q} .
\end{equation} 
Here $q = 1,2, \ldots$.

\section{Geodesics}

By keeping in mind the metric (\ref{b1.4}) 
we outline here certain useful relations for  geodesics  for a generic static  
spherically-symmetric metric written in Buchdal (quasi global) parametrization
\begin{equation} \label{b4.m}
 ds^2 =- A(r) dt ^2 + \frac{dr^2}{A(r)} + C(r) d \Omega^2,
\end{equation}
where $A = A(r) > 0$ is the red-shift function, $C(r) > 0$ is the central function, 
$d\Omega^2=d\theta^2+\sin^2\theta d\phi^2$ is the metric of the $2$-dimensional sphere.
 
\subsection{Geodesic equations}

Geodesic equations can be derived from the Lagrangian 
\beq \label{4.Lag}
    L = \frac{1}{2}  g_{\mu \nu}(x) \dot{x}^{\mu} \dot{x}^{\nu}
\eeq
by using the Euler-Lagrange equations
\beq \label{4.ELeqs}
    \frac{d}{d\tau}\left(\frac{\partial L}{\partial \dot{x}^{\mu}}\right)
     - \frac{\partial L}{\partial x^{\mu}} = 0,
\eeq
where $\dot{x}^{\mu}=dx^{\mu}/d\tau=u^{\mu}$ is 
the 4-velocity vector ($x^{\mu} = x^{\mu}(\tau)$, $\mu =0, 1, 2, 3$), 
 $\tau$ is the proper time for a massive point-like particle 
moving along a  timelike geodesic
and affine parameter in case of null  geodesic, correspondingly. 
We normalize 4-velocity vector  as following 
\beq \label{4.normal}
     g_{\mu \nu}(x)  u^{\mu} u^{\nu} = -k = 2 E_L = 2 L,
\eeq
where $k=-1,0,1$ for spacelike, null and timelike geodesics, respectively. 
The quantity $E_L = L$ 
is the energy integral of motion for the Lagrange equations, corresponding 
to Lagrangian (\ref{4.Lag}). The spacelike geodesics are irrelevant for our consideration,
so we put $k= 0,1$.

For the metric 
(\ref{b4.m})  the Lagrangian $L$ has the folllowing form
\beq \label{4.Lagrangian}
     L  = - A(r) \dot{t}^2 + (A(r))^{-1} \dot{r}^2 + C(r) (\dot{\theta}^2 + \sin^2 \theta  \dot{\phi}^2).
\eeq
 
Without loss of generality we consider null or timelike geodesics in the equatorial plane: 
\beq \label{4.theta} 
\theta = \pi/2.  \nonumber
\eeq
In this case the Lagrange equation in variable $\theta$ is satisfied identically. Other
Lagrange equations are governed by the reduced Lagrangian which  
corresponds to the $3D$ section of the metric 
 (\ref{b4.m}) 
\beq \label{4.Lagrangian1}
     L_{*}
    = - A(r) \dot{t}^2 + (A(r))^{-1} \dot{r}^2 + C(r) \dot{\phi}^2.
\eeq

For cyclic coordinates $t$ and $\phi$ we obtain 
the following integrals of motion  
\beq \label{4.ELhat}
  \varepsilon = A(r) \dot{t},\qquad  l = C(r) \dot{\phi}.
\eeq
They are associated for $k=1$ with the total energy 
$E= \varepsilon m$ and angular momentum $L= l m$ of   
a test (neutral point-like) particle of mass $m >0$. 

 Eq. (\ref{4.normal})  has the following form 
for the  line element  (4.1) 
\beq \label{eq:k}
    -  A(r) \dot{t}^2 + 
       (A(r))^{-1} \dot{r}^2 + C(r) \dot{\phi}^2 = -k. 
\eeq
By using Eqs. (\ref{4.ELhat}), we obtain the following differential equation: 
\beq \label{4.EqWithEL}
    -\frac{ \varepsilon^2}{A} + 
    \frac{ \dot{r}^2}{A} + \frac{l^2}{ C} = -k,
\eeq
which may be rewritten in terms of the effective potential $U$:
\beq \label{4.EqForR}
    \dot{r}^2 + U = \varepsilon^2,
\eeq
which is explicitly given by
\beq \label{4.eff_pot}
    U = U(r) = A(r) \left( k +  \frac{l^2}{ C(r) }   \right).   
    \eeq

One can readily verify that   the Lagrange equation  for the radial coordinate $r$ 
reads as follows
 \beq \label{4.EqForRtrue}
   2 \ddot{r} +  \frac{d U}{d r} = 0.
\eeq

For  $\dot{r} \neq 0$ it  can be also obtained just by differentiating Eq.
 (\ref{4.EqForR}) with respect to parameter $\tau$ and then dividing 
 the result by $\dot{r}$.  In case $\dot{r} = 0$ the radial equation reads
\beq \label{4.EqForR0}
    \frac{d U}{d r }= 0.
\eeq
It does not follow from Eq. (\ref{4.EqForR}) and should be considered separately.

\subsection{Radial null geodesics}

Consider the simplest radial trajectories of photons ($k=0$), corresponding 
to the case of zero angular momentum ($l=0$). In this case the effective potential
  \eqref{4.eff_pot} vanishes:
\begin{equation}\label{b.54}
k=0, \; l = 0 \seq U = 0.
\end{equation}

From the condition of constant value of $\theta$ (motion in the equatorial plane)
 and Eq. \eqref{4.ELhat}, we can write
\begin{equation}\label{b.55}
\phi = \phi_0 = \const,\quad \theta = \theta_0 = \frac{\pi}{2}
\end{equation}
(here $\theta_0$ can take any value) and
\begin{equation}
{\dot{r}}^2 = \varepsilon^2.\label{b.56}
\end{equation}

The line element (4.1) 
takes the form
\begin{equation} \label{b.57}
ds^2 = - A(r)dt^2 +  A(r)^{-1} dr^2 = 0,
\end{equation}
from which we obtain
\begin{equation} \label{b.58}
dt^2 = A(r)^{-2}dr^2 \seq dt =\pm A(r)^{-1} dr,
\end{equation}
and consequently
\begin{equation} \label{b.59}
t - t_0 = \pm \int_{r_0}^r A(y)^{-1} dy.
\end{equation}

Here, the ``$-$'' sign corresponds to the inward motion (toward the center). 
For  $r \to r_{h} = 2 \mu$ ($r_{h}$ is horizon radius) we obtain
\begin{equation} \label{b.61}
t \to + \infty,
\end{equation}
yielding the behaviour of the coordinate time near the horizon.

The ``$+$'' sign corresponds to the outward motion (away from the center).
If $r \to +\infty$, then
\begin{equation} \label{b.62}
t-t_0 \sim r - r_0. 
\end{equation}

From \eqref{b.56} we have
\begin{equation} \label{b.63}
\dot r^2 = \varepsilon^2 \seq \dot r = \pm \varepsilon \seq r - r_0 =  \pm \varepsilon (\tau - \tau_0).
\end{equation}

For the radial variable, we obtain a linear dependence on the affine parameter $\tau$ along the geodesic. This is a consequence of the Buchdahl parametrization chosen for the metric (4.1).

\subsection{Circular null geodesics}

Consider null geodesics with non-zero angular momentum
$l \neq 0$. In this case the effective potential \eqref{4.eff_pot} 
has the following form
\begin{equation}\label{Unull}
U(r) = A(r)\frac{l^2}{C(r)}.
\end{equation}
For the simplest case of circular orbits of a constant radius $r_*$
\begin{equation} \label{b.64}
r = r_* = \const, \quad l \neq 0,
\end{equation}
the trajectories form a sphere called a \textit{photon sphere}. 
It follows from \eqref{4.EqForR0},
\eqref{Unull}, and \eqref{b.64},
that the radius $r_*$ of the photon sphere can be determined from the relation
\begin{equation}\label{b.65}
\frac{d}{dr}\left( \frac{A}{C} \right) \biggr|_{r = r_*} = 0.
\end{equation}

From Eqs. \eqref{4.ELhat} it follows that
\begin{equation} \label{b.67}
A(r_*)\dot t = \varepsilon \seq t - t_0 = \frac{\varepsilon}{A(r_*)}(\tau - \tau_0),
\end{equation}
\begin{equation}\label{b.68}
C(r_*)\dot\phi = l \seq \phi - \phi_0 = \frac{l}{C(r_*)}(\tau - \tau_0),
\end{equation}
where $t_0,\tau_0,\phi_0$ are arbitrary constants. Thus,
\begin{equation}\label{b.69}
\phi - \phi_0 = \Omega(t - t_0),
\end{equation}
where
\begin{equation}\label{b.70}
\Omega = \frac{l A(r_*)}{\varepsilon C(r_*)}
\end{equation}
is the cyclic ``orbital'' frequency. Here and in what follows, we set for speed of light $c = 1$.

From Eq. \eqref{4.EqForR} we obtain:
\begin{equation}\label{b.71}
U(r_*) = \varepsilon^2 \seq \varepsilon =
\pm |l|\sqrt{\frac{A(r_*)}{C(r_*)}},
\end{equation}
and hence
\begin{equation}
\label{b.72}
\Omega = \frac{lA(r_*)}{\varepsilon C(r_*)} =
\pm \frac{l}{|l|}\sqrt{\frac{A(r_*)}{C(r_*)}}.
\end{equation}

 \subsection{Generic null non-circular and non-radial geodesics}


Consider non-radial, non-circular trajectories of photons in the equatorial plane 
$\theta = \pi/2$ satisfying for almost all
values of the affine parameter (except singular points):
\begin{equation}\label{b.81}
\dot r \neq 0,
\end{equation}
for which the radial equation \eqref{4.EqForRtrue} follows from \eqref{4.EqForR}, also
expressible as:
\begin{equation}\label{b.82}
\frac{dr}{d\tau} = \pm \sqrt{\varepsilon^2 - U(r)}.
\end{equation}

Using Eqs. \eqref{4.ELhat}, we derive the relations
\begin{equation}\label{b.83}
\frac{dt}{dr} = \pm \frac{\varepsilon}
{A(r)\sqrt{\varepsilon^2 - U(r)}},
\end{equation}
\begin{equation}\label{b.84}
\frac{d\phi}{dr} = \pm \frac{l}{C(r)\sqrt{\varepsilon^2 - U(r)}},
\end{equation}
whose integration, along with equation \eqref{b.82}, 
leads to the formal quadratures for solutions to the geodesic equations $(\dot r \neq 0)$:
\begin{equation}\label{b.85}
t - t_0 =  \pm \varepsilon\int_{r_0}^r\frac{d\bar r}{A\left( \bar r \right)\sqrt{\varepsilon^2 - U(\bar r)}},
\end{equation}
\begin{equation}\label{b.86}
\phi - \phi_0 =  \pm l\int_{r_0}^r\frac{d\bar r}{C(\bar r)\sqrt{\varepsilon^2 - U(\bar r)}},
\end{equation}\label{b.87}
\begin{equation}
\theta = \frac{\pi}{2},
\end{equation}
\begin{equation}\label{b.88}
\pm \int_{r_0}^r\frac{d\bar r}{\sqrt{\varepsilon^2 - U(\bar r)}} = \tau - \tau_0.
\end{equation}

\subsection{Spiral trajectory cases}

If we substitute into Eq. \eqref{b.86} the relation
\begin{equation}\label{b.90}
\varepsilon = \pm \sqrt{U(r_*)},
\end{equation}
which follows from \eqref{b.71} and corresponds to the circular motion, we obtain an infinite
spiral winding around the photon sphere with an infinite number of revolutions (a ``snail-like'' trajectory):
\begin{equation}\label{b.91}
\phi - \phi_0 =  \pm l\int_{r_0}^r\frac{d\bar r}{C(\bar r)\sqrt{U(r_*) - U(\bar r)}}.
\end{equation}
Express the potential \eqref{Unull} as follows:
\begin{equation}\label{b.92}
U(r) = l^2u(r),\quad u(r) := \frac{A(r)}{C(r)}.
\end{equation}

Then, relations \eqref{b.90} and \eqref{b.91}  for the spiral trajectory become:
\begin{equation} \label{b.93a}
\varepsilon = \pm |l|\sqrt{u(r_*)},
\end{equation}
\begin{equation}\label{b.93}
\phi - \phi_0 =  \pm \frac{l}{|l|}\int_{r_0}^r\frac{d\bar r}{C(\bar r)\sqrt{u(r_*) - u(\bar r)}}.
\end{equation}

To analyze the asymptotic behavior of the angular variable
as $r \to r_*$, expand $u(r)$ near $r_*$:
\begin{align}
&u(r) = u(r_*) + \frac{du}{dr}\biggr|_{r = r_*}( r - r_*) \nonumber
\\
&+\frac{1}{2} \frac{d^2u}{dr^2}\biggr|_{r = r_*} ( r - r_*)^2 + O\left(( r - r_*)^{3} \right). \label{b.94}
\end{align}
Assuming that, accordingly to \eqref{b.65}, $r_*$ is the point of maximum of the effective potential,
 we have the relations
\begin{equation}\label{b.95}
\frac{du}{dr}\biggr|_{r = r_*}\!\!\! = 0,\qquad \frac{d^2u}{dr^2} \biggr|_{r = r_*} \!\!\!= 2D < 0,
\end{equation}
from which we obtain:
\begin{equation}\label{b.96}
u(r) = u(r_*) - |D|( r - r_*)^2 + O\left((r - r_*)^3\right).
\end{equation}
Here 
\begin{equation}\label{b.96D}
D \equiv   \frac{1}{2}\frac{d^2u}{dr^2}\biggr|_{r = r_*} < 0
\end{equation}
and we deal with the case of instability of circular orbits.

Relation \eqref{b.93} describes two spiral trajectories:

(a) Outer spiral:
\begin{equation}\label{b.97}
r_* < r < r_0,
\end{equation}

(b) Inner spiral:
\begin{equation}\label{b.98}
{r}_* > r > r_0.
\end{equation}

Here, we consider the case (a), i.e. the outer spiral trajectory.
Setting (without loss of generality) $l > 0$, $\phi_0 = 0$, and choosing the ``$-$'' sign in
\eqref{b.93}, we obtain:
\begin{align}
\phi(r) &= - \int_{r_0}^r\frac{d\bar r}{C(\bar r)\sqrt{u(r_*)  - u(\bar r)}} \nonumber \\
           &= \int_r^{r_0}\frac{d\bar r}{C(\bar r)\sqrt{u(r_*) - u(\bar r)}}. \label{b.99}
\end{align}

Using the expansion (\ref{b.94}), we derive the asymptotic formula as
$r \to r_* + 0$ ($r_* < r < r_0$):
\begin{align}
\phi(r)&\sim\int_r^{r_0}\frac{d\bar r}{C(r_*)\sqrt{|D|(\bar r - r_*)^2}} \nonumber \\
 &=\int_r^{r_0}\frac{d\bar r}{C(r_*)(\bar r - r_* )\sqrt{|D|}} \nonumber \\
 &= \frac{1}{C(r_*)\sqrt{|D|}}\ln\frac{r_0 - r_*}{r - r_*}. \label{b.100}
\end{align}
This can be rewritten as:
\begin{equation}\label{b.101}
\exp\left( \beta_* \phi \right)\sim\frac{r_0 - r_*}{r - r_*}, \text{ as } r \to r_*,
\end{equation}
or, equivalently:
\begin{equation}\label{b.102}
(r_0 - r_*)\exp(- \beta_*\phi)\sim r - r_*,
\end{equation}
where parameter $\beta_*$ is defined as:
\begin{equation}\label{b.103}
\beta_{*} = C(r_*) \sqrt{|D|}.
\end{equation}

Fig. 1 illustrates the asymptotic parametrized curve \eqref{b.102} for a certain set of parameters 
$r_*$, $r_0$, and $\beta_*$.

\begin{figure}    
\centering     
\includegraphics[width=\linewidth]{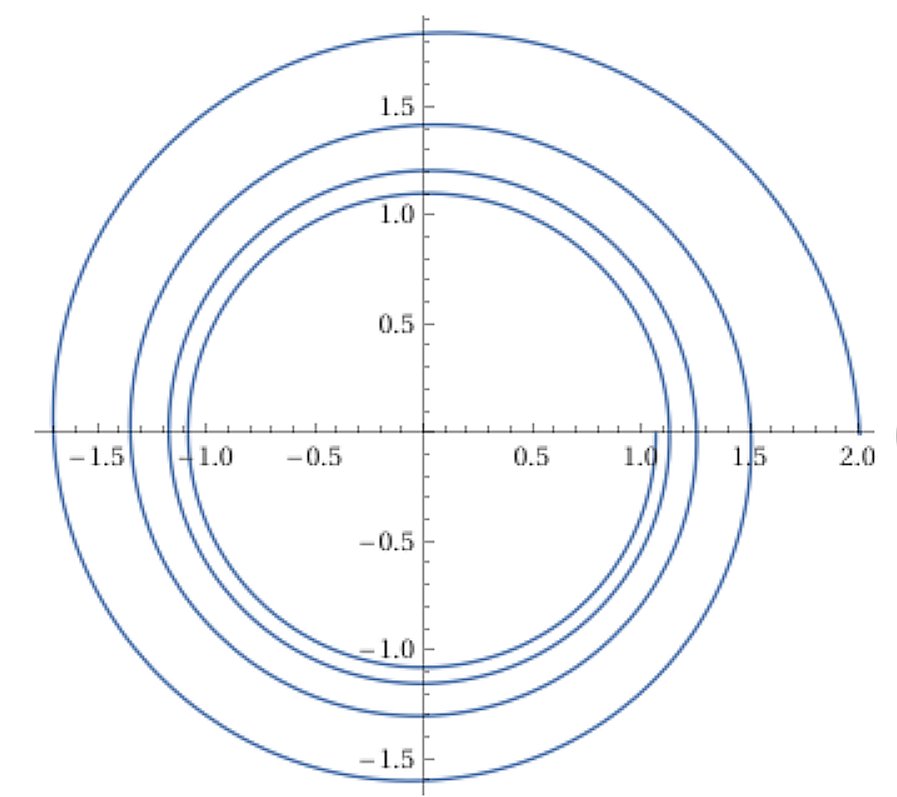}
\caption{\small
	The asymptotic parametrized curve \eqref{b.102} for $r_* = 1$, $r_0 = 2$,
	and $b_* = 0,1103\cong \ln 2/(2\pi)$, with the
	angle $\phi$ varying from $0$ to $8\pi$.} \label{fig1}
\end{figure}         

}

\section{Photon spheres}

Here we deal with the photon spheres for the metric 
\eqref{b1.4} written in Buchdal form (\ref{b4.m}),
where $A(r)$ and $C(r)$ according to Eq. (\ref{b1.4}) can be written as
\begin{eqnarray} \label{b4.A}
 A(r)&=&A=H^{-a}\left( 1-\frac{2\mu}{r} \right), \\
 C(r)&=&C=H^{a} r^2, \quad a = 2/q,    \label{b4.C}
\end{eqnarray}
where  
 \begin{eqnarray}
 \nonumber 
 H(r) = 1+\frac{P}{2\mu} \left[ 1-\left(1-\frac{2\mu}{r}\right)^q \right] \\
  = 1 + p (1 - z^q)  \label{4.H}
  \end{eqnarray}
is the moduli function, $\mu  > 0$, $P > 0$, $p = P/(2 \mu)$, $q = 1,2, \dots$. 
Here and in what follows we use $z$-variable: 
\begin{equation}
 \label{b4.Hz}
  z = 1-\frac{2\mu}{r},  \qquad  r = \frac{2\mu}{1- z}.
  \end{equation}
We note that $0 < z < 1$ for $r > 2\mu$.

\subsection{Master equation}

For the effective potential  $U = U(r) = \frac{l^2 A(r)}{C(r)} $
we get 
\begin{eqnarray} 
U= \frac{l^2 H^{-2a}( 1-\frac{2\mu}{r})}{r^2} \label{b4.5U} \\
 = \frac{l^2 z (1-z)^2(1 + p (1-z^q))^{-2a}}{(2\mu)^2}.
                     \label{b4.5Vf}
\end{eqnarray}
From the extremum condition   
\begin{equation} \label{b4.6V}
 \frac{d U}{dr} = 0
\end{equation}
for $l^2  \neq 0$, we obtain 
\begin{equation} \label{b4.6VV}
\frac{1}{U} \frac{d  U}{ dz}  
 = \frac{1}{z} - \frac{2}{1 -z}  + \frac{4 p z^{q-1} }{1 + p (1-z^q)}  =0,
\end{equation}
and hence we get the master equation \cite{BI} 
\begin{equation} \label{b4.6m}
  p z^{q+ 1} - 3 p z^{q} + (1+ p)(3z - 1) = 0,
  \end{equation}
which gives the radius of photonic sphere (see (\ref{b4.Hz})).

{ \bf Proposition \cite{BI}}. {\em For any $p > 0$, $\mu > 0$ and $q \in {\bf N}$,  the 
master equation (\ref{b4.6m}) has only one solution
obeying $0 < z < 1$, which corresponds the extremum point of the effective potential $U(r)$. 
}

In other words the Proposition states the existence and uniqueness of photon 
sphere (outside the event horizon) for the metric (\ref{b1.4}) 
for any proper set of parameters.  

The proposition  is proved in Ref. \cite{BI}. We denote this 
point of extremum by $r_*$.  Thus, in term of the variable $z$ we get
that the point  $z_* = 1- 2\mu/r_*$ is a unique solution
to Eq.  (\ref{b4.6m}) for  $z \in (0,1)$.

The maximum  of the effective potential thus becomes
\begin{equation}
   U_*= U(r_*) = \frac{l^2}{H^{2a}(r_*) r_*^2}
    \left(1-\frac{2\mu}{r_*}\right).   \label{b4.9V}
\end{equation}

The second derivative with respect to radial coordinate in the point of extremum is given by 
the following relation \cite{BI}
\begin{equation}
 \frac{d^2 U}{dr^2}\bigg|_{r=r_*}
    =  - \frac{1}{2}  \left(\frac{2\mu}{r_{*}^2}\right)^2  U_*  \mathcal{B}(z_*),  \label{b4.V2}
\end{equation}
 where
\begin{equation}
  \mathcal{B}(z) = \frac{3}{2} q - \frac{2 (q - 2 ) z}{(1-z)^2}. \label{b4.B}
\end{equation}
It was proved in Ref. \cite{BI} that $ \mathcal{B}(z_*) > 0$ and 
hence
\begin{equation}
  \frac{d^2 U}{dr^2}\bigg|_{r=r_*} = 2 D l^2 < 0, \label{b4.B0}
\end{equation}
 for all $q \in {\bf N}$.

Now we consider  three cases $q = 1, 2, 3$, when the master equation 
(\ref{b4.6m}) may be solved in radicals for all values of $p > 0$.  
We also explore the limiting case $q = + \infty$.

\subsection{The case $q = 1$}
 Let us consider the case  $q = 1$ ($a = 2)$.
 In this case the master equation (\ref{b4.6m}) is just a quadratic one 
 with the only one root: 
 \begin{equation} \label{b4.6zpm}
   z_* = z_*(1,p) = \frac{-3 + \sqrt{4 p(p+1)+9}}{2p},
   \end{equation}
  belonging to interval $(0,1)$. This root corresponds
  to the radius of photon sphere $r_* = \frac{2\mu}{1- z_*} > 2 \mu$. 

 For all values $p > 0$ we have \cite{BI}
  \begin{equation} \label{b4.6bz0}
     1/3 < z_{*}(1,p) < 1.
  \end{equation}

 We note that for $q = 1$ our  metric (\ref{b1.4}) 
 is coinciding with the Reissner-Nordstr\"om metric  
    \begin{equation} 
   ds^2 =   - \bar{f}(\bar{r}) dt^2 +   
   (\bar{f}(\bar{r}))^{-1} d\bar{r}^2 + \bar{r}^2  d \Omega^2,
           \label{b4.12RN}
  \end{equation}
 where $\bar{r}= r + P$, $\bar{f}(\bar{r}) = 1 - \frac{2GM}{\bar{r}} + \frac{Q^2}{2 \bar{r}^2}$, 
 with $Q^2 = 2P (P + 2 \mu)$.  
 Here $\bar{r}_* = r_* + P$ corresponds to the position of the unstable, 
circular photon orbit in the Reissner-Nordstr\"om spacetime.

\subsection{The case $q = 2$}
 
 Now we put $q=2$ ($a = 1$). The master equation (\ref{b4.6m})
 in this case is just cubic one. It has a unique (real) solution 
 $z_* = z_*(2,p)$ for any $p >0$ belonging to interval $(1/3,1)$
  \begin{equation} \label{b4.13z0}
    z_* = z_*(2,p) = Z^{1/3} - p^{-1} Z^{-1/3} +1,
  \end{equation}
where
 \begin{equation} \label{b4.13Z}
     Z = Z(p) = \frac{1}{p} \left( \sqrt{ 1 + \frac{1}{p}} - 1 \right).
   \end{equation}

Here for all  $p > 0$ we get  $1/3 < z_*(2,p) < 1$ \cite{BI}.

\subsection{The case $q = 3$}

Let us consider the last case $q=3$, when the master equation (\ref{b4.6m})
is of fourth power and has a solution in radicals \cite{BI}:
 \begin{eqnarray} 
       z_* = z_*(3,p) = \frac{1}{2} \sqrt{X} - \frac{\sqrt{Y}}{4\sqrt{3}} + \frac{3}{4}, 
     \quad  \label{b4.14z0} \\
      X = - \frac{3 \sqrt{3}}{2} \left(1 -  \frac{8}{p}\right)  Y^{-1/2} - Z^{1/3} \nonumber \\ 
              -  \frac{5(p+1)}{3 p} Z^{- 1/3} +  \frac{9}{2}.  \quad  \label{b4.14X} \\
     Y = 12 Z^{1/3} + 27 + 20 \left(1 + \frac{1}{p} \right) Z^{- 1/3}, \quad  \label{b4.14Y}  \\ 
     Z = \frac{p+1}{2 p^2} ( 9 + 3^{-3/2} \sqrt{2187 - 500 p (p+1)}  ). \quad \label{b4.14Z} 
  \end{eqnarray}

It may be verified  that $z_* = z_*(3,p)$, given by relations (\ref{b4.14z0})-(\ref{b4.14X}), 
is real for all $p > 0$ and obey the inequalities $1/3 < z_* < \frac{3 - \sqrt{5}}{2} \approx 0,382$ \cite{BI}.

\subsection{The case $q = + \infty$}

Let us consider the limiting case $q \to + \infty$, when 
$z^q \to 0$ ($0 < z < 1$). In this case thee master equation (\ref{b4.6m}) reads 
as  $3z -1 = 0$.   We get a root $z_* = 1/3$  
and hence  $r_* = 3 \mu = 3GM$ ($c =1$). Thus, we are led   
to the radius of  photon sphere for  Schwarzschild black hole 
the metric 
  $ds^2= \frac{dr^2}{1-\frac{2\mu}{r}} + r^2 d\Omega^2  
   - \left(1-\frac{2\mu}{r}\right) dt^2$ which appears
   in the limiting case (as $q \to + \infty$) of our metric (\ref{b1.4}).

\section{Some applications}

In this section we present two examles of applications for photon sphere solutions:
related to shadow angles and eikonal quasinormal modes.
In both examples the ``rescaled'' effective potential $u = u(r) = U(r)/l^2$
is used  
\begin{equation} 
u(r) = \frac{ (H(r))^{-2a}( 1-\frac{2\mu}{r})}{r^2} . \label{b4.5u} 
 \end{equation}

\subsection{Black hole shadows}

Consider relation \eqref{b.99} for the special solution describing a spiral-like geodesic curve with 
an infinite ``winding'' angle. This curve acts as a boundary between two classes of solutions and corresponds to
the critical angle $\vartheta_{sh}\in (0, \pi/2)$, which is the
angle between the tangent to the curve \eqref{b.99} at the point $(r_0, \phi_0)$ and the radial line.

If a light ray is emitted from $(r_0, \phi_0)$ at an angle smaller than the critical angle:
\begin{equation}\label{b.104}
\vartheta < \vartheta_{sh},
\end{equation}
the ray will cross the photon sphere and, in the case of a black hole, enter the event horizon and fall into the black hole.
If the emission angle satisfies
\begin{equation}\label{b.105}
\vartheta > \vartheta_{sh},
\end{equation}
the ray will not reach the photon sphere and will escape to infinity after a finite number of revolutions.

Since $r_*$ is the point of maximum of the effective potential, we have:
\begin{equation}\label{b.107}
u(r_*) > u(r) > 0.
\end{equation}
for all $r > r_*$.
The \textbf{shadow
angle} $\vartheta_{sh}$ can be calculated using the
differential relation derived from \eqref{b.99},
\begin{equation}\label{b.108}
 d\phi =  - \frac{dr}{C(r)\sqrt{u(r_*) - u(r)}},
\end{equation}
and the $2D$ section of the metric (\ref{b4.m}) having the form 
\begin{equation}\label{b.109}
\!\! dl^2 = B(r)dr^2 + C(r)d\phi^2, \;\; B(r)=A(r)^{-1}.
\end{equation}

Using \eqref{b.108} and \eqref{b.109}, we obtain:
\begin{align}
&\tan\vartheta_{sh} = \frac{\sqrt{g_{\phi \phi}(r_0)}}{\sqrt{g_{rr}(r_0)}}
 \left| \frac{d\phi}{dr} \right|_{r = r_0} \nonumber \\
&= \frac{\sqrt{C(r_0)}}{\sqrt{B(r_0)}}\frac{1}{C(r_0)\sqrt{u(r_*) - u\left( r_0 \right)}}  
   \nonumber\\
 &= \frac{\sqrt{A(r_0)}}{\sqrt{C(r_0)}}\frac{1}{\sqrt{u(r_*) - u(r_0)}}
  = \frac{\sqrt{u(r_0)}}{\sqrt{u(r_*) - u\left( r_0 \right)}} \nonumber \\
 &= \frac{1}{\sqrt{\frac{u(r_*)}{u\left( r_0 \right)} - 1}}. \label{b.110}
\end{align}
Using the trigonometric identity we get
\begin{equation}\label{b.112}
\sin^2\vartheta_{sh} = \frac{1}{1 + \cot^2\vartheta_{sh}} = \frac{u(r_0)}{u(r_*)},
\end{equation}
or, equivalently \cite{PTs}:
\begin{equation}\label{b.113}
\sin\vartheta_{sh} = \sqrt{\frac{u(r_0)}{u(r_*)}}.
\end{equation}
Therefore, taking into account Eq.~\eqref{b.107}, we get for the shadow angle:
\begin{equation}\label{b.114}
\vartheta_{sh} = \arcsin\sqrt{\frac{u(r_0)}{u(r_*)}}, \qquad 0<\vartheta_{sh}<\frac{\pi}{2},
\end{equation}
for all $r_0 > r_*$.
Here, $r_0$ describes the radial coordinate (position) of a point light source or light receiver (observer),
and $r_*$ is the radius of the photon sphere.

For big enough values of (observer's) radial cooordinate $r_0$ obeying $r_0 \gg \mu$ and 
$r_0 \gg P$ the shadow angle may be found (approximately, with certain accuracy) 
by using  the following asymptotical formula:
   \begin{equation}\label{b.115}
   \vartheta_{sh} = \frac{b_*}{r_0 } + O \left( \frac{1}{r_0^2}  \right), 
   \end{equation}
as $r_0 \to + \infty$. Here
 \begin{equation}\label{b.116}
   b_{*} = \frac{1}{\sqrt{u(r_*)}} 
   \end{equation} 
is critical impact parameter
\cite{PTs}.

\subsection{Quasinormal modes }

The radius of  photon sphere  and the effective
potential (\ref{4.eff_pot}) $U(r) = U(r, l^2 )$
 ($k =0$) with ``quantized''  $l^2$,
i.e. replaced by $l (l+1)$,
 \begin{equation} 
      \label{5.1}
   V_{eik} (r) = \frac{A(r)l (l+1)}{C(r)} , 
 \end{equation}
 where $l = 0, 1, 2, \dots $,
may be used for calculation of the ``spectrum'' of  quasinormal modes (QNM) for 
certain test fields in the eikonal approximation \cite{KSt}, i.e. when $l \to + \infty$.
The functions $A(r)$ and $C(r)$ are defined by  formulas (\ref{b4.A}), (\ref{b4.C}), respectively.

Here we outline this ``eikonal'' spectrum for a neutral massless scalar (test) field
defined in the background given
by the metric (\ref{b1.4}) with parameters $P > 0$ and $\mu > 0$. 

The scalar field satisfies standard Klein-Fock-Gordon equation
\begin{equation}
\label{5.QNM2}
\Delta \Psi =\frac{1}{\sqrt{\vert g \vert }}\partial_{\mu}
 (\sqrt{\vert g\vert }g^{\mu \nu } \partial_{\nu } \Psi )=0.
\end{equation}

We are seeking the solution to equation (\ref{5.QNM2}) by using a  separation of variables 
\begin{equation}
\label{5.QNM3}
\Psi = e^{-i \omega t} \Psi_{1}(r) Y_{lm},
\end{equation}
where $Y_{lm}$ are  spherical harmonics and $\Psi_{1}(r)$ obey  
QNM boundary conditions with $|\Psi_{1}(r)| \to + \infty$ in two limits:
  as $r \to 2\mu$ and $r \to + \infty$.

By choosing an appropriate
sign for  $\omega$ we get the asymptotic relations (as $l \to + \infty$)
on real and imaginary parts of complex  $\omega$ in the
eikonal approximation \cite{BI}

\begin{eqnarray}
  {\rm Re}(\omega)  &=&   \left(l + \frac{1}{2}\right) H_{*}^{-a} r_*^{-1} z_*^{1/2},  \label{b4.11Re}
      \\
 {\rm Im}(\omega)  &=&  - \left(n + \frac{1}{2}\right) 
 H_{*}^{-a} \mu r_{*}^{-2} \mathcal{B}_{*}^{1/2}, \label{b4.11Im}
\end{eqnarray}
where $H_{*} = H (r_{*})$,  $r_{*} =  2\mu /(1 - z_{*})$, and  $z_{*} \in (0,1)$ is solution 
to master equation (\ref{b4.6m}) and $\mathcal{B}_* = \mathcal{B}(z_*)$, where 
$\mathcal{B}(z)$ is defined in (\ref{b4.B}).
(In (\ref{b4.11Re}) and  (\ref{b4.11Im}) the $O(1/(l + 1/2))$ terms are omitted.) 

Here we confirm that  the parameters of the unstable circular null geodesics
around  stationary spherically symmetric and asymptotically flat black holes are 
in correspondence with the eikonal part of  quasinormal modes of these black holes.
See \cite{b40} and references therein. Indeed, due to (\ref{b.72}) 
and  (\ref{b4.11Re}) we get up to $O(1/(l + 1/2))$ term
\begin{equation}
\label{5.Omega}
 {\rm Re}(\omega) = \left(l + \frac{1}{2}\right) |\Omega|,
\end{equation}
where $\Omega$ is cyclic  frequency given by (\ref{b.72}),
corresponding to ``orbital photon rotation'' at photon sphere. 

\section{Conclusions}

In this paper we have  explored null geodesics for non-extremal black hole  solutions in  $4d$ gravitational 
model with  anisitropic fluid from Refs. \cite{DIM,BI}. The black hole solutions are defined by  moduli function $H(r)$,
which depend upon parameters: $P > 0$, $\mu > 0$ and $q = 1, 2, \dots$,  where $2 \mu$ is horizon radius.

We have studied the  circular null geodesics with constant radius $r = r_*$ 
($r_*$ is the radius of photon  sphere), which are governed by $(q+1)$-order polynomial  master equation. 
For any choice of $P > 0$, $ \mu  > 0$ and  $q = 1, 2, \dots$ this equation has one and only one 
solution which corresponds to circular photon orbit with radius $r_* > 2 \mu$. Any such orbit 
is unstable.

 For any $q = 1, 2, \dots$ we have also explored relations for black hole shadow angles and 
 presented the spectrum of  quasinormal modes  for  a  test neutral massless scalar field 
 in the eikonal approximation.
 
 Here we have restricted ourselves to the case of light-like (null) geodesics. The next step of research may be concerned  with the study of time-like geodesics (e.g. ISCO) describing the motion  of massive point-like particles
 in the background of the black hole solutions with anisotropic fluid. This and some other issues (e.g. related to quasinormal modes)  will be addressed in forthcoming publications. Another topic of interest may be related to higher dimensional  extension of considered $4d$ solutions.       

\vspace{5pt}

{\bf FUNDING}

For V.D.I. and S.V.B. this research was funded by RUDN University, scientific project number FSSF-2023-0003. 

\newpage

\vspace{5pt}



\end{document}